\input harvmac
\input epsf

\def\figin{\epsfcheck\figin}\def\figins{\epsfcheck\figins}
\def\epsfcheck{\ifx\epsfbox\UnDeFiNeD
\message{(NO epsf.tex, FIGURES WILL BE IGNORED)}
\gdef\figin##1{\vskip2in}\gdef\figins##1{\hskip.5in}
\else\message{(FIGURES WILL BE INCLUDED)}%
\gdef\figin##1{##1}\gdef\figins##1{##1}\fi}
\def\DefWarn#1{}
\def\figinsert{\goodbreak\topinsert}
\def\ifig#1#2#3#4{\DefWarn#1\xdef#1{fig.~\the\figno}
\writedef{#1\leftbracket fig.\noexpand~\the\figno}%
\figinsert\figin{\centerline{\epsfxsize=#3mm \epsfbox{#2}}}
\bigskip\medskip\centerline{\vbox{\baselineskip12pt
\advance\hsize by -1truein\noindent\footnotefont{\sl Fig.~\the\figno:}\sl\ #4}}
\bigskip\endinsert\noindent\global\advance\figno by1}

\def\R{{\bf R}}
\def\Z{{\bf Z}}

\def\b{\beta}

\def\l{\lambda}

\def\cE{{\cal E}}

\def\({\left(}
\def\){\right)}
\def\<{\left\langle}
\def\>{\right\rangle}

\def\frac#1#2{{#1 \over #2}}

\def\sqr#1#2{{
{\vbox{\hrule height.#2pt
\hbox{\vrule width.#2pt height#1pt \kern#1pt
\vrule width.#2pt} \hrule height.#2pt}}}}

\lref\wit{
E.~Witten, ``Topological sigma models,'' Commun.\ Math.\ Phys.\ {\bf
118}, 411 (1988); ``Mirror manifolds and topological field theory,''
arXiv:hep-th/9112056.  }

\lref\katz{
D.A.~Cox and S.~Katz, {\it Mirror Symmetry and Algebraic Geometry},
Mathematical Surveys and Monographs, AMS, Providence, 1999.
}

\lref\mirb{
{\it Mirror Symmetry}, Clay Mathematics Monographs, Vol. 1, C.~Vafa
and E.~Zaslow, eds., AMS-CMI, Providence, 2003.  
}

\lref\kkv{
S.~Katz, A.~Klemm and C.~Vafa, ``M-theory, topological strings and
  spinning black holes,'' Adv.\ Theor.\ Math.\ Phys.\ {\bf 3}, 1445
(1999) [arXiv:hep-th/9910181].  }

\lref\gova{
R.~Gopakumar and C.~Vafa, ``M-theory and topological strings. I, II''
  arXiv:hep-th/9809187, arXiv:hep-th/9812127.}

\lref\osv{
H.~Ooguri, A.~Strominger and C.~Vafa, ``Black hole attractors and the
topological string,'' Phys.\ Rev.\ D {\bf 70}, 106007 (2004)
  [arXiv:hep-th/0405146].
}

\lref\andy{
D.~Gaiotto, A.~Strominger and X.~Yin, ``New connections between 4D and
5D black holes,'' arXiv:hep-th/0503217.}

\lref\ionv{
A.~Iqbal, N.~Nekrasov, A.~Okounkov and C.~Vafa, ``Quantum foam and
topological strings,'' hep-th/0312022.}  

\lref\nv{
A.~Neitzke and C.~Vafa, ``Topological strings and their physical
applications,'' arXiv:hep-th/0410178.  }

\lref\bcov{
M.~Bershadsky, S.~Cecotti, H.~Ooguri and C.~Vafa, ``Kodaira-Spencer
theory of gravity and exact results for quantum string amplitudes,''
Commun.\ Math.\ Phys.\ {\bf 165}, 311 (1994),
 [arXiv:hep-th/9309140].}

\lref\narain{
I.~Antoniadis, E.~Gava, K.S.~Narain and T.R.~Taylor, ``Topological
amplitudes in string theory,'' Nucl.\ Phys.\ B\ {\bf 413}, 162 (1994)
[arXiv:hep-th/9307158].  }

\lref\mnop{
D. Maulik, N. Nekrasov, A. Okounkov and R. Pandharipande, "Gromov-Witten
theory and Donaldson-Thomas theory, I-II", math.AG/0312059;
math.AG/0406092.  }

\lref\orv{
A.~Okounkov, N.~Reshetikhin and C.~Vafa,
``Quantum Calabi-Yau and classical crystals,'' hep-th/0309208.}

\lref\DT{
S.K.~Donaldson and R.P.~Thomas, ``Gauge theory in higher dimensions,'' in
{\it The Geometric Universe; Science, Geometry, And The Work Of Roger
Penrose}, Oxford University Press, 1998.}

\lref\dvv{
R.~Dijkgraaf, E.~Verlinde and M.~Vonk,
{``On the partition sum of the NS five-brane,''} hep-th/0205281.} 

\lref\vw{
C.~Vafa and E.~Witten,
  ``A Strong coupling test of S duality,''
  Nucl.\ Phys.\ B {\bf 431}, 3 (1994)
  [arXiv:hep-th/9408074].
}

\lref\tobe{
M.~Cheng, R.~Dijkgraaf, J.~Manschot and E.~Verlinde, work in progress, to
appear.}

\lref\nov{
N.~Nekrasov, H.~Ooguri and C.~Vafa,
  ``S-duality and topological strings,''
  JHEP {\bf 0410}, 009 (2004)
  [arXiv:hep-th/0403167].
}

\lref\dmmv{
R.~Dijkgraaf, J.~M.~Maldacena, G.~W.~Moore and E.~P.~Verlinde,
  ``A black hole Farey tail,''
  arXiv:hep-th/0005003.
}

\lref\gsy{
D.~Gaiotto, A.~Strominger and X.~Yin,
`` From AdS3/CFT2 to Black Holes/Topological Strings,''
hep-th/0602046.
}


\Title{\vbox{\baselineskip11pt 
\hbox{HUTP-06/A001}
\hbox{ITFA-2006-05} }} 
{\vbox{ 
\centerline{M-theory and a Topological String Duality
} }}
\centerline{ 
Robbert Dijkgraaf,$\!{}^{1,2}$ Cumrun Vafa$^3$ and Erik Verlinde$^1$}
\medskip
\medskip
\medskip
\vskip 8pt
\centerline{\it $^1$Institute for Theoretical Physics, and $\,\,{}^2$KdV
 Institute for Mathematics,}
\centerline{\it University of Amsterdam, Valckenierstraat 65,
1018 XE Amsterdam, The Netherlands.}
\medskip
\centerline{\it
$^3$Jefferson Physical Laboratory, Harvard University, Cambridge, MA
02138, USA.}
\medskip
\medskip
\medskip
\noindent 
We show how the topological string partition function, which is known
to capture the degeneracies of a gas of BPS spinning M2-branes in
M-theory compactified to 5 dimensions, is related to a 4-dimensional
D-brane system that consists of single D6-brane bound to
lower-dimensional branes. This system is described by a topologically
twisted $U(1)$ gauge theory, that has been conjecturally identified
with quantum foam models and topological strings.  This also explains,
assuming the identification of Donaldson-Thomas invariants with this
$U(1)$ gauge theory, the conjectural relation between DT invariants
and topological strings. Our results provide further mathematical
evidence for the recently found connection between 4d and 5d black
holes.

\medskip

\Date{February 2006}

\newsec{Introduction}

A-model topological string \wit\ has been well studied from physical
and mathematical perspectives (see \refs{\katz,\mirb} for a detailed
review of the subject as well as \nv\ for a more recent update).  It
is a source of inspiration for quite a number of diverse areas on the
physical and mathematical fronts.  On the physics side it captures,
worldsheet instantons for sigma models, relevant configurations for
superpotential corrections for superstrings as well as counting the
BPS black hole degeneracies for 5d black holes \gova\ and more
recently connected to melting of crystals as a model for quantum
gravitational foam and a $U(1)$ topologically twisted Yang-Mills
theory on the Calabi-Yau \ionv .  On the mathematical front it is
known as the study of Gromov-Witten invariant, which has been defined
rigorously and studied extensively.  It has also recently been
conjecturally connected to Donaldson-Thomas (DT) invariants \mnop.

The relation between the physics and mathematical developments has
been basically clear: The worldsheet instanton corrections captured by
A-model topological strings is simply what one means by Gromov-Witten
invariant computations.  The quantum gravitational foam captured by
the $U(1)$ topologically twisted Yang-Mills theory, is believed
(though not proven) to be the physical formulation of Donaldson-Thomas
invariants.  So to complete the missing link we have to connect the 5d
BPS spinning black hole degeneracies with the $U(1)$ topologically
twisted Yang-Mills theory on the Calabi-Yau.  This would be relating
what is called the Gopakumar-Vafa (GV) invariants, with the DT invariants.
It is the aim of this brief note to explain this link from an M-theory
perspective.

The idea is rather simple: the topological string partition function
has an interpretation in M-theory, where it can be reformulated in
terms of the integer GV invariants \gova, that calculate the
BPS-degeneracies of M2-branes with angular momentum wrapped on
2-cycles in the Calabi-Yau.  

On the other hand, the DT invariants, or the partition function of the
(presumably equivalent) $U(1)$ topologically twisted super-Yang-Mills
theory, compute the bound states of wrapped D2 and D0-branes to a
single D6-brane, that completely covers the Calabi-Yau. To relate the
two set of invariants we make use of the recent observation of \andy\
that connects the physics of 4d and 5d black holes.

The configuration of D-branes relevant for the DT calculation can be
lifted to M-theory as follows. In M-theory the D2-branes simply become
M2-branes. The D6-brane located at a fixed point in the spatial $\R^3$
is represented by a Taub-NUT geometry. Asymptotically this looks like
$\R^3 \times S^1$, where the circle is the extra 11${}^{\rm th}$
dimension. At the center the geometry is smooth and given by
$\R^4$. The D0-brange charge is associated with a $U(1)$ isometry of
the Taub-Nut. At infinity this is the momentum along the circle, but
at the center it becomes a rotation. Therefore, the D-brane system can
be identified with a collection of spinning M2-branes in the TN
geometry.

In this paper we will argue that the degeneracy of the D-brane bound
state, which by a most reasonable conjecture is computed by the DT
invariant, is equal to the degeneracy of a free gas of M2-branes.  So,
in this way are able to relate the second quantized Hilbert space of
5d BPS states with the DT invariants and thereby complete the circle
of ideas.

The plan for this paper is as follows: In section 2 we review the
relation between A-model topological strings and BPS degeneracies for
5d black holes.  In section 3 we review the conjectures relating
topological strings with the $U(1)$ topologically twisted gauge
theory, which is believed to be the physical formulation of DT
invariants.  In section 4 we connect these two viewpoints by
considering an extra D6-brane wrapping the Calabi-Yau and thus derive
the conjectural relation between $U(1)$ topologically twisted gauge
theory ({\it i.e.} conjecturally DT invariants) and the topological
A-model.

\newsec{A-model topological string and 5d spinning black holes}

It has been known that topological strings compute certain amplitudes
in compactifications of superstrings on Calabi-Yau manifolds down to
four dimensions \refs{\bcov,\narain}.  In particular
the genus $g$ amplitude $F_g$ compute corrections to the ${\cal N}=2$
effective action of the form
\eqn\tops{\int d^4x \, d^4\theta \ F_g(t) {\cal W}^{2g-2}, }
where $t$ are the K\"ahler moduli and $\cal W$ is the graviphoton
superfield.  By viewing these corrections as the contribution of
D2 and D0-branes, and interpreting this in the context of
embedding IIA theory in M-theory, it was argued in \gova\ that
$F_g$ encodes the degeneracy of BPS spinning 5d black holes
corresponding to M2-branes wrapped over 2-cycles of the
Calabi-Yau. In particular, it was argued that the Kahler moduli
dependence of topological string captures the charge dependence of
these BPS states and the string coupling constant dependence of
the amplitudes captures the spin content of the BPS states.

The starting point of the GV calculation \gova\ is a compactification
of M-theory on a Calabi-Yau 3-fold $X$ down to 5 dimensions. One is
interested in computing a thermal partition function of BPS-states and
hence the time is Euclidean and compactified on a $S^1$. The idea
behind the GV calculation is to identify this thermal circle with the
11${}^{th}$ M-theory direction. The 5D angular momentum couples to the
field strength $F_{gph}$ of the graviphoton field.  Since the
counting of BPS-states corresponds to an index, the result is
independent of the period $\beta$ of the Euclidean time-circle, that
plays the role of the IIA string coupling constant. In this way we can
relate the perturbative picture at $\b =0$ to the strong coupling
limit $\b = \infty$.

In the limit $\beta\to 0$ the M2-branes are mapped on to IIA
fundamental strings. Due to the presence of the graviphoton field
strength the string world sheet calculation reduces precisely to
topological string theory, and counts the contribution of world sheet
instantons. The topological string coupling constant is equal to the
combination 
$$
\lambda=\beta  F_{gph},
$$ 
which is kept fixed in the $\beta \to 0$ limit. Hence the
topological string partition function that computes the BPS quantities
receives contributions of all genera and even non-perturbative
effects.

Alternatively, in the $\beta\to\infty$ limit, the light objects are
the D2 and D0-branes, and their bound states.  The free energy
$F(t,\lambda)$ of the type IIA string can thus be rewritten as a sum
of contributions of a Schwinger-type calculation for each of these
D2-D0 bound states.  Here the D0-brane charge corresponds to the
momentum along the Euclidean time-circle, and hence has no direct
physical interpretation from the four-dimensional space-time
perspective. The computation of \gova\ is based on integrating out
these charged particles that become light in the strong coupling
limit, where the type IIA theory should be reformulated in
M-theory. Therefore this Schwinger calculation essentially takes place
in 5 dimensions. A note of caution that we will return to: the
D0-brane charge that is being summed over in these virtual loops has
an interpretation as the momentum along the thermal circle. The
degeneracies are independent of this D0-brane charge, which should
{\it not} be confused with the spin of the M2-branes.

\subsec{Five-dimensional BPS degeneracies}

Consider a compactification of M-theory on a Calabi-Yau threefold $X$
down to 5 dimensions.  We are interested in the BPS states in this
theory that are charged under the $U(1)$'s and correspond to M2-branes
wrapped over two-cycles of the Calabi-Yau.  Let
$$Q\in H_2(X,{\bf Z})$$
denote a generic charge of such an M2-brane.  Such states will also
carry spin.  The spatial rotation group in 5 dimensions is
$SO(4)=SU(2)_L\times SU(2)_R$.  We can label a BPS state, in addition
to its charge $Q$, by its spin.  Let $(m_L,m_R)$ denote the spin
content $(2j^3_L,2 j^3_R)$ of the highest spin state in a given BPS
multiplet.  Let $N_Q^{(m_L,m_R)}$ denote the number of such BPS
states.  Define
\eqn\NQm{
N_Q^m=\sum_{m_R}(-1)^{m_R} N_Q^{(m_L,m_R)}.
}
This quantity is an invariant in the sense that, if one changes the
hypermultiplet moduli, this index does not change. The numbers
$N_Q^{(m_L,m_R)}$ can only change if the left and right spin contents
pair up, which leaves this index invariant since it only counts the
net number.  Note that $N_Q^m$ may be a positive or a negative
integer. As discussed in \kkv\ the spin content of an M2-bound state is
related to the topology of the holomorphic curve in the CY. A single
M2-brane wrapping a curve of genus $r$ gives a particular (reducible)
representation with maximal spin $r$. For our purpose it is useful to
concentrate only on the $j_L^3$ quantum number. So, in principle the
numbers $N_Q^m$ can receive contributions of curves of arbitrary high
genus. Nevertheless, we will assume that $N_Q^m$ are finite integers.

Now we are ready to rewrite the topological string amplitudes in terms
of $N_Q^m$ following \gova, see also \nv.  The topological string
partition function, that has a perturbative expansion of the form
\eqn\topstring{
Z(\lambda,t) = \exp \sum_{g\geq 0} \lambda^{2g-2} F_g(t),
}
depends on the Kahler moduli $t \in H^2(X,{\bf R})$ as well as on the
topological string coupling constant $\lambda$.  Then the result of
\gova\ is that this partition function has an alternative form
\eqn\mafo{
Z(\lambda, t)=\prod_{Q,m} \Bigl[\prod_{n\geq 0} (1-e^{\l(n+m)+ t \cdot Q})^n
\Bigr]^{N_Q^m}
}
It has the following interpretation in terms of a gas of M2-branes in
five dimensions: First, one notices that, after substituting 
relation \NQm\ in the above formula, each spin content
$m_L,m_R$ contributes a factor
$$
\prod_{n_1,n_2\geq 0} \left(
1-e^{\l_1(m_1+n_1+\half) }  e^{\l_2(m_2+n_2+\half)+ 
t \cdot Q} \right)^{\pm 1}.
$$
Here the quantum numbers $m_1,m_2$ are given by
$$
m_1 = {1\over 2}(m_L + m_R),\qquad
m_2 = {1\over 2}(m_L-m_R),
$$
and we introduced additional parameters $\l_1,\l_2$ that are eventually set to
$\l_1=\l_2=\l$. The parameter $t$ acts as a potential that couples to the
charge $Q$ of the M2-brane.  The $\pm 1$ in the exponent depends on
whether the (top element) of the BPS state is a boson or
fermion. Tracing back the derivation, one finds that the quantum
numbers $m_1,m_2$ refer to the intrinsic spins of the BPS-state.  
The numbers $n_1,n_2$ indicate the orbital helicities in each of the
two planes of ${\bf R}^4$.

More precisely, for each BPS state we get a field $\Phi(z_1,z_2)$
where $z_1,z_2$ denote the coordinates of ${\bf R}^4={\bf C}^2$.
Because of the BPS condition this field has a holomorphic expansion
$$
\Phi(z_1,z_2)=\sum_{n_1,n_2} \alpha_{n_1,n_2} z_1^{n_1}z_2^{n_2}.
$$
These holomorphic wave functions are the familiar lowest Landau levels
that describe a charged particle in a constant self-dual graviphoton
field. These wave functions are normalizable, since the proper
normalization includes a Gaussian factor. Therefore, these wave
functions are localized in the $\R^4$ plane.

Each factor in the above partition function can now be viewed as
counting the second quantized states of the corresponding quantum
field $\Phi$, where the oscillators $\alpha_{n_1,n_2}$ could be
bosonic or fermionic, depending on the internal spin of the field
$\Phi$.

\newsec{D-brane bound states and DT invariants}

Let us now turn to the counting of D-brane states in IIA string
theory.  Consider a single D6-brane wrapped on the Calabi-Yau manifold
$X$.  On the brane there is a maximally supersymmetric $6+1$
dimensional $U(1)$ gauge theory, that is naturally topologically
twisted. This gauge theory has non-trivial topological sectors that
correspond to bound states with lower-dimensional branes.

The Witten index of this field theory,
$$
{\rm Tr}\left[(-1)^F e^{-\beta H} \right],
$$
is the partition function on $S^1 \times X$ and, within a given
topological sector, this should computes the number of D-brane bound
states. Mathematicians have considered the corresponding moduli spaces
and the associated invariants (roughly the Euler numbers of these
moduli spaces) are known as Donaldson-Thomas invariants \DT. It is
conjectured in \ionv\ that the quantum gauge theory exactly computes these
mathematical invariants.

By taking the $\beta \to 0$ limit of the index we obtain a
six-dimensional $U(1)$ topological field theory, that was conjectured
to be the same as the topological string partition function
\refs{\ionv,\mnop}.  This second conjecture was based on the discovery
of a deep relationship between the statistical mechanical model of
crystal melting and topological string amplitudes \orv . Notice that,
since we shrink the thermal circle, it is most natural to perform a
$T$-duality along that circle. In this way we obtain immediately a
Euclidean D5-brane wrapping the CY, whose world-volume theory gives
the six-dimensional abelian gauge theory that we consider here. (For
the case of the B-model, the relation with the world-volume theory of
the 5-brane was also made in \dvv.)

By considering a non-trivial gauge bundle, the topological field
theory on the D6-brane naturally includes bound states with lower
dimensional branes. We will 
label the charges of the D6-D4-D2-D0 system as 
$$ 
(p_0,p,q,q_0) \in H^0(X) \oplus H^2(X) \oplus H^4(X) \oplus H^6(X). 
$$ 
Mathematically, the possible D-brane charges are
given by the Chern classes of the corresponding gauge bundle $\cE$. In
the case of a $U(1)$ theory the bundle $\cE$ is of rank one and
necessarily has to be an ideal sheaf. The physical charges are given
by the Chern characters
$$
ch_k(\cE) = {1\over k! (2\pi)^k} {\rm Tr} \,F^k 
\in H^{ev}(X,{\bf Q}).
$$
These Chern characters can be expressed in terms of the integer Chern
classes $c_i(\cE) \in H^{ev}(X,\Z)$. For the case of a single D6-brane
with $ch_0(\cE)=rk(\cE)=1$ the charges are given by
\eqn\charges{
\eqalign{
D6:\ \ & p_0 = rk =1,  \cr
D4:\ \ & p = c_1,  \cr
D2:\ \ & q = ch_2 = -c_2 + {1\over 2} c_1^2 - {1\over 48} p_1(X), \cr
D0:\ \ & q_0 = 2 ch_3 = c_3 - c_1 c_2 + {1\over 3} c_1^3
- {1\over 24} p_1(X) c_1. \cr
}}
Here we have also included the gravitational correction coming from
the first Pontryagin class $p_1(X) \sim {\rm Tr}\, R^2$ of the CY
space.

In the absense of D4-branes, the coupling to the D2 and D0-branes is
given by the action
$$
S =\lambda \int_X c_3(F)+\int k\wedge ch_2(F)
$$
where $[k]= t$ gives the Kahler class of the CY and $\l$ will be
identified with the topological string coupling. 
The resulting partition function has an expansion in terms of the integer
Donaldson-Thomas invariants $DT_{q,q_0}$
$$
Z(\lambda,t) = \sum_{q,q_0} DT_{q,q_0} \ e^{\l q_0 +t \cdot q}
$$
The conjecture of \refs{\ionv,\mnop} is that this equals the topological
string partition \topstring. The aim of this paper is to explain this
relation by connecting directly the D-brane state counting to the
5-dimensional BPS counting of \gova.  An alternative physical
explanation for this conjecture, based on quantum gravitational foam
for the A-model topological string, was given in \ionv. This independent
explanation provides additional support for the arguments presented in the
following section.

Although the DT invariants are usually defined in the case $c_1=0$,
{\it i.e.} without D4-branes, they can in principle also be determined
for the case $c_1 \not = 0$. This is a rather trivial generalization,
because a non-trivial $c_1$ does not change the nature of the
singularities of the bundle. This flux is carried by smooth line
bundles with non-trivial topologies, so intuitively it can be thought
of as being smeared out over the CY space. The invariants only depends
on the moduli space of the singularities, which are captured by the
Chern classes $c_2$ and $c_3$. So, turning on a D4-brane magnetic
charge
$$
p=c_1(\cE) \in H^4(X,\Z).
$$ 
does not change the value of the invariant, but it merely changes the
definition of what we call D2 and D0-brane charges $(q,q_0)$, as we 
can read off from formula \charges,
\eqn\chargeshift{
\eqalign{
q & \to q + {1\over 2} p^2,\cr
q_0 & \to q_0 - q\cdot p - {1\over 6} p^3 -{1\over 48} p_1(X) p. \cr
}}
The fact that the D-brane degeneracy is independent of this shift,
can also be obtained by duality arguments \andy.

\newsec{D-brane degeneracies from a gas of M2-branes}

As we explained in section 2.1, the topological string partition
function has an interpretation in terms of a gas of five-dimensional
black holes. We now like to argue that this same partition function
also gives the degeneracies of the four-dimensional D-brane system of
section 3, and hence is related to Donaldson-Thomas
theory. Furthermore, instead of counting a gas of 5d BPS states, it
enumerates the states of a single BPS object in four dimensions.

\subsec{Lift to M-theory}

Let us start with the IIA D-brane system of the previous section, without
any D4-branes for the moment. The thermal partition function of this
system is computed in the ten-dimensional background
\eqn\bg{
X \times \R^3 \times S^1,
}
with $X$ the Calabi-Yau three-fold and $S^1$ the thermal circle of
radius $\b$. In 
the lift to M-theory the single D6-brane gets represented by a 
$k=1$ Taub-NUT geometry. So, the eleven-dimensional background is
$$
X \times TN \times S^1,
$$
The D2 and D0-branes lift to M2-branes and KK-momenta.

This is precisely the configuration used in \andy\ to connect
4d and 5d BPS computations.
The Taub-NUT space can be considered as an $S^1$ fibration over $\R^3$,
where the circle shrinks to zero at the origin (the location of the
D6-brane) and attains a finite
radius $R$ at infinity. The metric can be written with $\chi \in S^1$,
and ${\vec x}\in \R^3$, as 
\eqn\tnmetric{
 ds^2_{TN} = R^2 \left[ {1\over V}(d\chi
+
\vec{A} \cdot d{\vec x})^2 + V d{\vec x}^2 \right],
}
with
$$
V =  {1 + {1\over |{\vec x}|}}, \qquad \vec{\nabla}\times
\vec{A} = \vec{\nabla} V.
$$ 

Since we now have an M-theory compactification with two (asymptotic)
circles, we can take either one of these as the eleventh
direction. The two inequivalent reductions to IIA string theory give
rise to the two interpretations of the partition function that we want
to relate in this paper. The first reduction along the TN circle gives
back the starting configuration of D-branes in the background \bg. The
second reduction uses the thermal circle and gives a $X \times TN$
geometry. We will argue that this leads to the GV interpretation of
the topological string partition function. 

Note that the exchange of the two circles can be done entirely within
the IIA theory. It is the well-known ``9-11 flip'' that is obtained as
the duality transformation $T\cdot S \cdot T$, with $T$ being the
$T$-duality on the remaining $S^1$ and $S$ the strong-weak coupling
duality of the IIB string. Under this transformation the D6-brane in
our starting configuration gets mapped to the TN geometry as
follows. The first $T$-duality turns it into a Euclidean D5 wrapping
the CY. The $S$-duality makes this into a NS5-brane, which then by the
second (transversal) $T$-duality becomes the Taub-NUT. Similarly, the
D2-branes become Euclidean D1-branes and subsequently fundamental
strings.  Finally, the D0-branes become momentum modes along the $S^1$
of the TN fibration.

The TN metric is a hyper-K\"ahler geometry with a $U(1) \times
SO(3)$ isometry group, that acts in the obvious way on the variables
$\chi$ and ${\vec x}$. Near the origin the geometry is smooth, and
locally given by $\R^4$.  This is most easily seen by writing the
metric on $\R^3$ in spherical coordinates $(r,\theta,\phi)$.  After
making the substitution $r=\rho^2$ and using that near the origin $V
\sim \rho^{-2}$ one recognizes the flat $\R^4$-metric in spherical
coordinates $(\rho,\chi,\theta,\phi)$, where the angular part is
represented by the Hopf fibration of $S^3$ over $S^2$: here $\chi$ is
the coordinate along the fiber, while $\theta$ and $\phi$ parametrize
the base. The isometries of the Taub-NUT geometry leave the origin
invariant and act as $SO(4)=SU(2)_L\times SU(2)_R$ rotations in
$\R^4$. The $SO(3)$ rotations of the $S^2$-base correspond to
$SU(2)_R$, while the $U(1)$ that shifts $\chi$ is contained in
$SU(2)_L$.  This in particular implies that the $U(1)$ quantum number
is given by an angular momentum quantum number at the origin. This
suggest the following re-interpretation of the D0-brane quantum
number. It equals the KK momentum at infinity, but in the interior it
is identified with the {\it total} angular momentum. This receives
contributions from both the intrinsic spin $m$ of the M2-branes and
from their orbital angular momentum $n$. This identification of the
4d D0-brane charge with the 5d angular momentum was one of the
essential observations of \andy.

\subsec{Physical derivation}

\ifig\tn{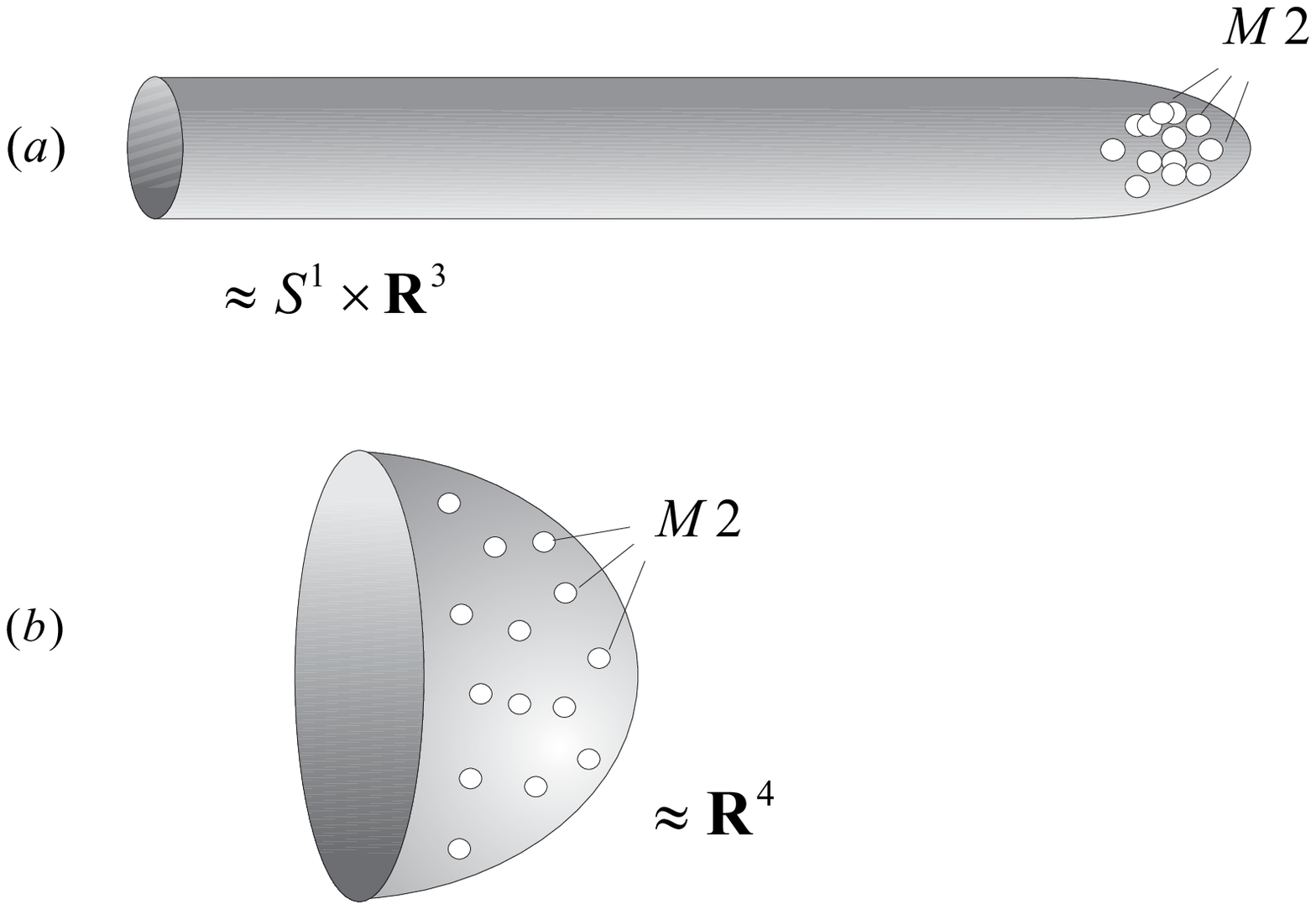}{100}{
The small and large radius limits of the Taub-NUT geometry interpolate
between $(a)$ a bound state of 4d D-branes and $(b)$ a free gas of spinning
5d M2-branes.}

From an M-theory perspective the perturbative gauge theory description
of the D-brane system naturally arises in the limit of small TN radius
$R$.  In the $R\to 0$ limit the M2-branes are clustered together in
the origin and form a single bound state from the IIA point of view as
illustrated in \tn$(a)$. This is in accordance with the well-known fact
that codimension-four branes (in this case D6 and D2) can form a bound
state.

The idea is now to consider the opposite limit of large radius $R$,
where we make use of the fact that the M-theory partition function
with supersymmetric boundary conditions is independent of $R$. As we
can see from the metric \tnmetric, changing the radius is equivalent
to rescaling the entire geometry. So, we are effectively zooming in to
the center of the Taub-NUT, see \tn$(b)$. In the $R\to\infty$ limit
the M2-branes disperse and form a free gas of charged, spinning
BPS-particles in $\R^4$. This is where we make contact with the GV
computation.

We are now in a position to give a four-dimensional interpretation of
the GV partition function. First of all, the D2-charge $q$ of the 4d
bound state is given by the sum of the charges $Q_i$ of the individual
M2-branes in the 5d gas
$$
q = \sum_i Q_i.
$$
Similarly, the D0-brane charge $q_0$ is equal to sum of the individual
angular momenta $\ell_i$, which for each M2-brane is the sum of the
internal and orbital spin quantum numbers $m_i$ and $n_i$ 
$$
q_0 =  \sum_i \ell_i,\qquad \ell_i = m_i + n_i.
$$
Therefore, the GV partition function \mafo, when interpreted as
counting the D-brane degeneracies, should be written as
\eqn\dbrane{
Z(\lambda, t)=\prod_{Q,\ell} (1-e^{\l \ell + t \cdot
Q})^{c_Q^\ell} }
where the multiplicities
$$
c_Q^\ell=\sum_{n=0}^\ell n N_Q^{\ell-n}
$$
can in some way be regarded as counting the number of ``irreducible''
D2-D0 states inside the D6. In fact, this formula is a generalization
of a similar result obtained for the counting of D4-D0 bound states on
a 4-manifold $M$. As has been shown in \vw\ for a single
D4-brane with $N$ D0-branes the number of bound states is given by the
Euler number of the instanton moduli space (or more precisely in this
case the Hilbert scheme of points on $M$). These
degeneracies $d_N$ are encoded in the generating function
$$
\sum_N d_N e^{tN} = \prod_{k>0}\left(1 - e^{tk}\right)^{-\chi(M)} 
$$
which can be obtained as a special case of \dbrane\ for the case
of $X = M \times T^2$.

As explained, the D-brane partition function should also be captured
by the Donaldson-Thomas theory. Combining the above ingredients, we
see that the conjectured relation between the GV-invariants and the
DT-invariants is a direct consequence of the connection between 4d and
5d black holes of \andy. Conversely, the independent mathematical
evidence for the relation between these two mathematical invariants
should be regarded as supporting the 4d-5d connection.

\subsec{Inclusion of D4-branes}

We have seen in section 3 it is very natural to include also D4-branes
in the gauge theory. When we appropriately shift the charges $q$ and
$q_0$ this will not change the degeneracies. This is also clear from
the M-theory perspective. Here the D4-branes will be represented as
M5-branes that wrap a 4-cycle in the CY space. They are therefore
string-like objects in the Taub-NUT geometry that wind around the
$S^1$ fibers. Since that circle is contractible in the core of the TN,
these M5-branes are not supported by the topology and can be
contracted to zero size leaving behind a flux for the 4-form field
strength $G$. In the scaling limit where we zoom in to the interior of
the Taub-NUT geometry, this flux becomes diluted and has no effect on
the microscopic counting of the black hole states. However, the
definition of the asymptotic charges $q$ and $q_0$ gets modified, due
to the presence of the Chern-Simons term in the M-theory effective
action. This analysis has been performed in \andy, where exactly the 
relations \chargeshift\ were found. It is gratifying that there is
such a direct relation between the quantities in the gauge theory and
the effective supergravity. Perhaps this observation can eventually
be extended to a full gauge theory/gravity correspondence.

\newsec{Additional Remarks}

The connection between the topological string and the counting of
D-brane bound states that we have discussed here is different from
that of the OSV conjecture \osv\ . First of all, the present relation
is exact, but holds only for a single D6-brane. On the other hand, in
the OSV conjecture the relation with the topological string only
arises asymptotically in the limit for large magnetic charges, {\it
i.e.} D4 and D6. This regime is outside the scope of this present
paper. Furthermore, in the OSV conjecture the square of the
topological string partition function appears. Finally, the role of
the topological string coupling constant $\l$ and moduli $t$ is
different. 

Our M-theory derivation of the topological string duality makes use of
two different reductions to the IIA string: one along the circle in
Taub-NUT, one along the thermal circle. The GV relation with the
topological string \gova\ arises from the latter reduction. In this
paper the first was used to obtain the D-brane system. But also in
this case a role is played by the topological string, namely along the
lines of the OSV conjecture. Hence, these two occurrences of the
topological string are related by the $TST$ duality chain that
interchanges the two circles. This role of S-duality has been
anticipated in \refs{\nov,\osv} and was a crucial ingredient in the
recent paper \gsy. 

To use our exact counting formula for the computation of the entropy
of 4d black holes, we have to extend the present analysis from the
case of a single D6-brane to a large number of D6-branes. One expects
that the full answer is invariant under the $S$-duality and that
leading growth of the number of states is captured by the OSV formula,
that appears after the duality map, very much like the Cardy formula
in conformal field theory. This picture is very analogous to the
``black hole Farey tail'' of \dmmv, and is also suggested by the
analysis of the $(0,4)$ CFT in \gsy. Semi-classically, the two
appearances of the topological string are separated by a
Hawking-Page-type phase transition in the dual gravitational system
from a thermal gas to a black hole \tobe.
 
\medskip 
\centerline{\bf Acknowledgments} 

We would like to thank M. Cheng, J. Manschot, A. Strominger for
fruitful discussions. R.D. and E.V. wish to thank the Harvard Physics
Department for kind hospitality. The research of R.D. and E.V. was
supported by a NWO Spinoza grant and the FOM program {\it String
Theory and Quantum Gravity}.  The research of C.V. was supported in
part by NSF grants PHY-0244821 and DMS-0244464.
 
\listrefs

\end